\documentclass[12pt]{article}

\usepackage{graphicx}

\usepackage{cite}
\usepackage[colorlinks]{hyperref}

\hypersetup{
    linkcolor=red,          
    citecolor=black,        
    urlcolor=blue           
}

\usepackage{lineno} 

\usepackage{setspace}

\usepackage{amssymb,amsfonts,amsmath}
\usepackage{lscape}

\usepackage[margin=1in]{geometry}

\usepackage[labelfont=bf,labelsep=period,justification=raggedright]{caption}

\date{}

\usepackage{lscape}
\usepackage{soul}
\usepackage{lineno}

\begin{document}
\linenumbers

\begin{flushleft}
{\Large
\textbf{A phylogenomic perspective on the radiation of ray-finned fishes based upon targeted sequencing of ultraconserved elements}
}
\medskip
\medskip
\bigskip

Michael E. Alfaro$^{1,2,\ast}$, 
Brant C. Faircloth$^{1,2}$,
Laurie Sorenson$^{1}$,
Francesco Santini$^{1}$
\\
\vspace{0.25in}
\footnotesize{\textsuperscript{1}Department of Ecology and Evolutionary Biology, University of California, Los Angeles, CA, USA}\\
\footnotesize{\textsuperscript{2}These authors contributed equally to this work}\\
\vspace{0.25in}
$\ast$ E-mail: michaelalfaro@ucla.edu
\end{flushleft}

\doublespacing

\newpage
\section*{Summary}
Ray-finned fishes constitute the dominant radiation of vertebrates with over 30,000 species. Although molecular phylogenetics has begun to disentangle major evolutionary relationships within this vast section of the Tree of Life, there is no widely available approach for efficiently collecting phylogenomic data within fishes, leaving much of the enormous potential of massively parallel sequencing technologies for resolving major radiations in ray-finned fishes unrealized. Here, we provide a genomic perspective on longstanding questions regarding the diversification of major groups of ray-finned fishes through targeted enrichment of ultraconserved nuclear DNA elements (UCEs) and their flanking sequence. Our workflow efficiently and economically generates data sets that are orders of magnitude larger than those produced by traditional approaches and is well-suited to working with museum specimens. Analysis of the UCE data set recovers a well-supported phylogeny at both shallow and deep time-scales that supports a monophyletic relationship between \emph{Amia} and \emph{Lepisosteus} (Holostei) and reveals elopomorphs and then osteoglossomorphs to be the earliest diverging teleost lineages. Divergence time estimation based upon 14 fossil calibrations reveals that crown teleosts appeared ~270 Ma at the end of the Permian and that elopomorphs, osteoglossomorphs, ostarioclupeomorphs, and euteleosts diverged from one another by 205 Ma during the Triassic. Our approach additionally reveals that sequence capture of UCE regions and their flanking sequence offers enormous potential for resolving phylogenetic relationships within ray-finned fishes.

\newpage

\section*{Introduction}
The ray-finned fishes (Actinopterygii) constitute the dominant radiation of vertebrates on the planet including more than 32,000 species and equaling or exceeding richness estimates for the combined total of birds, mammals, and reptiles. Despite a long history of systematic study, resolution of phylogenetic relationships within this vast radiation remains elusive.  Studies based upon traditional morphological and single-gene, PCR-based molecular approaches have succeeded in delineating several major lineages of ray-finned fishes, but conflict over how these lineages are related to one another remains. For example, the earliest morphological studies of ray-finned fishes unite gar (\emph{Lepisosteus}) with the bowfin (\emph{Amia}) in the clade Holostei \cite{Nelson1969, Jessen1972} though this clade is not recovered in some later analyses  \cite{Olsen1984, Patterson1973}. The early branching of teleost lineages has also been historically contentious.  Systematists agree on the four earliest-diverging lineages: the osteoglossomorphs (bony-tongues; arawanas, elephant fishes, and allies), the elopomorphs (tarpons, bonefishes, and eels), the ostarioclupeomorphs (anchovies and herrings, minnows, characins, catfishes, and electric eels), and the euteleosts (salmons, pikes, lizardfishes, and perch-like fishes). However, there is disagreement over both the relationships among these groups and the basal divergences within euteleosts. Recent morphological and molecular studies support a sister-group relationship between ostarioclupeomorphs and euteleosts \cite{Arratia2001, Li2008, Inoue2003}, but beyond this there is little agreement regarding the relationship among these ancient teleost lineages. Morphological analyses alternatively place the osteoglossomorphs \cite{Patterson1977} or the elopomorphs \cite{Arratia2001, Arratia2010, Arratia2004, Cloutier2004} as the sister group to all other teleosts and the remaining lineages sister to the ostarioclupeomorph/euteleost clade. Some molecular analyses place elopomorphs and osteoglossomorphs as the sister group to remaining teleosts \cite{Le1993, Broughton2010} while others recover a basal divergence between osteoglossomorphs and other teleosts \cite{Inoue2003,  Inoue2005}.

Recently, Near \emph{et al.} \cite{Inoue2005} used wide-spread taxonomic sampling, in conjunction with sequence collected from nine commonly used nuclear genes, to provide a more comprehensive phylogenetic hypothesis of relationships among fishes. Their results supported the monophyly of the Holostei, suggested that the elopomorphs formed the earliest diverging teleost lineage\cite{Near2012}, and provided a new timescale for the divergence of ray-finned fishes. Although promising, these new insights into the radiation of actinopteryigians relied upon a relatively modest number of genomic markers, and the stability and timing of these relationship encoded throughout the genomes of the target groups remain largely untested. One exception to this statement includes a recent study by Zou \emph{et al.} \cite{Zou2012} which used transcriptome sequences to examine basal divergences within euteleosts. However, the Zou \emph{et al.} \cite{Zou2012} study did not include several anciently diverging lineages (e.g. \emph{Amia}, osteoglossomorphs) informing questions about the early evolution of major groups of ray-finned fishes.

Phylogenomics and next-generation sequencing technologies offer enormous promise for resolving relationships within actinopterygians and other major sections of the Tree of Life. However, revolutions within genomics and informatics have had a surprisingly modest effect on data collection practices within the phylogenetics community:  most studies of non-model organisms continue to rely upon direct sequencing of a moderate number of loci, and workflows that do take advantage of massively parallel sequencing platforms remain bottlenecked by cross-species amplification of phylogenetically informative loci.  Several alternatives to traditional phylogenetic workflows exist that help to overcome the inefficiencies of gene-based sequencing.  One class of these methods is exemplified by the recent work of Zou \emph{et al.} \cite{Zou2012}, who used a combination of \emph{de novo} transcriptome sequencing, existing transcript data, and computational methods to identify 274 orthologous groups from which they inferred the phylogeny of the Actinopterygii. The benefits of their approach include the use of existing, transcript-related data sets (ESTs in GenBank); reasonably well-established data generation methods; and the collection of data from hundreds of loci across the genomes of the focal taxa.  Limitations of this approach include reliance on sampling fresh or properly preserved tissues (generally precluding the use of thousands of existing museum samples), dependence of the approach on expression patterns of the tissue sampled, and collection of data from fewer genomic locations than alternative methodologies.

A second class of phylogenomic methods involves sequence capture of nuclear regions flanking and including ultraconserved elements (UCEs) \cite{Faircloth2012}. Rather than sequencing expressed portions of the genome, the UCE-based approach involves enriching organismal DNA libraries for hundreds to thousands of UCEs and their flanking regions; sequencing these libraries using massively parallel sequencing; and assembling, aligning, and analyzing the resulting data using informatic tools. This approach has been successfully used in mammals \cite{McCormack2012a}, birds \cite{Faircloth2012, McCormack2012b}, and reptiles \cite{Crawfordetal2012} to generate phylogenomic data sets that contain at least one order of magnitude more characters than those generated using PCR and to resolve historically contentious sections of the Tree of Life \cite{McCormack2012a, Crawfordetal2012}.  The UCE approach differs from transcript-based phylogenomic studies \cite{Zou2012} because data collection is independent of expression pattern, researchers can prepare and enrich libraries from existing tissue collections, and UCE loci may be better conserved and more numerous across distantly related taxa \cite{McCormack2012a}.

Here, we apply the UCE approach to ray-finned fishes by developing a novel set of sequence capture probes targeting almost 500 UCE regions in ray-finned fishes. We use the UCE data to provide the first phylogenomic perspective based upon widespread sampling of hundreds of markers across the genome on long-standing controversies regarding relationships at the base of the ray-finned fish Tree of Life. These include whether \emph{Lepisosteus} and \emph{Amia} form a monophyletic group (the Holostei \cite{Nelson1969, Jessen1972, Grande2010}) and how the major lineages of teleosts, which constitute \textgreater 99\% of ray-finned fishes, are related to one another\cite{Arratia2001,  Arratia2010, Arratia2004, Cloutier2004,  DePinna1996, Li2008, Inoue2003, Zaragueta2002}. We also use 14 fossil calibrations to provide the first time-scale for ray-finned fishes based upon UCE regions and their flanking sequence. Our results reveal that sequence capture of UCE regions can efficiently and economically generate massive data sets with strong resolving power at both deep and shallow phylogenetic scales within fishes.

\section*{Materials and Methods}
\subsection*{Identification of UCE regions}
To identify ultraconserved elements (UCEs) in fishes, we used genome-to-genome alignments of stickleback (\emph{Gasterosteus aculeatus}) to medaka (\emph{Oryzias latipes}) to locate nuclear DNA regions of 100\% conservation greater than 80 bp in length. To enable efficient capture-probe design, we buffered these regions to 180 bp (where needed) by including equal amounts of medaka sequence 5' and 3' to each UCE. We aligned or re-aligned these buffered regions to the genome-enabled fishes (zebrafish, \emph{Danio rerio}, stickleback, medaka, and two species of puffers, \emph{Tetraodon nigroviridis} and \emph{Takifugu rubripes}) using LASTZ \cite{Harris2007}, keeping only non-duplicate matches of $\geq$ 120 bp and $\geq$ 80\% sequence identity across all species in the set. Based on the intersection of UCE loci across all fishes that were greater than 10 Kbp apart, we designed a pilot set of 120 bp sequence capture probes for each of the UCEs present among all members of the set by tiling probes at 4X density. We had these probes commercially synthesized into a custom SureSelect target enrichment kit (Agilent, Inc.). We used a higher than normal \cite{Tewhey2009} tiling density to help ameliorate potential sequence differences among species introduced by buffering shorter UCEs to 180 bp.

\subsection*{Library preparation, UCE enrichment, sequencing, and assembly}
We prepared DNA libraries from 18 fish species, including representatives of five acanthomorph orders and two families of perciforms (Table \ref{tab:read-stats}), by slightly modifying the Nextera library preparation protocol for solution-based target enrichment \cite{Faircloth2012} and increasing the number of PCR cycles following the tagmentation reaction to 20. Following library preparation, we substituted a blocking mix of 500 $\mu$M (each) oligos composed of the forward and reverse complements of the Nextera adapters for the Agilent-provided adapter blocking mix (Block \#3). We incubated species-specific libraries with synthetic RNA probes from the SureSelect kit for 24 h at 65$^{\circ}$C. We followed the standard SureSelect protocol to enrich DNA libraries following hybridization; we eluted clean, enriched DNA in 30 $\mu$L of nuclease free water; and we used 15 $\mu$L of enriched template in a 50 $\mu$L PCR reaction of 20 cycles combining forward, reverse, and indexing primers with Nextera polymerase to add a custom set of 24 indexing adapters \cite{FairclothandGlenn2012}. PCR clean-up was completed using Agencourt AMPure XP. We quantified enriched, indexed libraries using qPCR (Kapa Biosystems), and we prepared two library pools containing 10 libraries at equimolar ratios prior to sequencing. 

We sequenced each pool of enriched DNA using two lanes of a single-end 100 bp Illumina Genome Analyser (GAIIx) run. After sequencing, we trimmed adapter contamination, low quality bases, and sequences containing ambiguous base calls using a pipeline we constructed (\url{https://github.com/faircloth-lab/illumiprocessor}). We assembled reads, on a species-by-species basis, into contigs using Velvet \cite{Zerbino2008} and VelvetOptimiser. Following assembly, we used a software package (\url{https://github.com/faircloth-lab/phyluce}) containing a custom Python program (match\_contigs\_to\_probes.py) integrating LASTZ \cite{Harris2007} to align species-specific contigs to the set of probes/UCEs we used for enrichment while removing reciprocal and non-reciprocal duplicate hits from the data set. During matching, this program creates a relational database of matches to UCE loci by taxon. This program also has the ability to include UCE loci drawn from existing genome sequences, for the primary purpose of including available data from genome-enabled taxa as outgroups or to extend taxonomic sampling. We used this feature to include UCE loci we identified in the genome sequences of \emph{Gasterosteus aculeatus}, \emph{Haplochromis burtoni}, \emph{Neolamprologus brichardi}, \emph{Oreochromis niloticus}, \emph{Oryzias latipes}, \emph{Pundamilia nyererei}, \emph{Takifugu rubripes}, \emph{Tetraodon nigroviridis}, \emph{Gadus morhua}, and \emph{Lepisosteus oculatus}. After generating the relational database of matches to enriched sequences and genome-enabled taxa, we used additional components of PHYLUCE (get\_match\_counts.py) to query the database and generate fasta files for the UCE loci we identified across all taxa. Then, we used a custom python program (seqcap\_align\_2.py) to align contigs with MAFFT \cite{Katoh2005} and trim contigs representing UCEs, in parallel, across the selected taxa  prior to phylogenetic analysis \cite{Faircloth2012}. 

\subsection*{Phylogenetic Analyses}
The large number of UCE loci we collected create a vast potential space for partitioning data that makes a traditional evaluation of alternative partitioning strategies computationally intractable. As a result, we modeled nucleotide substitutions across the concatenated data set using two approaches.  For Bayesian analysis, we used a custom script (run\_mraic.py) wrapping a modified MrAIC 1.4.4 \cite{Nylander 2004} to find the best-fitting model for each UCE locus, we grouped loci having similar substitution models (selected by AICc) into the same partition, and we assigned the partition specific substitution model to all loci concatenated within each partition.  For maximum likelihood analyses, we maintained the partitions identified in the Bayesian analysis and we modeled each partition using the GTR+CAT approximation.  We performed Bayesian analysis of the concatenated data set using MrBayes 3.1 \cite{Ronquist2003} and two independent runs (4 chains each) of 5,000,000 iterations each, sampling trees every 500 iterations, to yield a total of 10,000 trees. We sampled the last 5,000 trees after checking results for convergence by visualizing the log of posterior probability within and between the independent runs for each analysis, ensuring the average standard deviation of split frequencies was $<$ 0.001, and ensuring the potential scale reduction factor for estimated parameters was approximately 1.0. We performed maximum likelihood analysis of the concatenated data in RAxML \cite{Stamatakis2008} using the rapid bootstrapping algorithm and 500 bootstrap replicates. 

Gene tree-species tree methods enjoy some advantages over the analysis of concatenated data sets under certain conditions \cite{Kubatko2007, Edwards2007, Edwards2009} but may also be sensitive to missing data \cite{Bayzid2012} and to the resolution of individual gene trees \cite{Castillo2010}. To minimize the number of  unresolved  gene tree topologies and maximize the number of topologies that overlapped in sampling the base of the actinopterygian tree, we selected a subset of the UCE contigs containing complete data for \emph{Polypterus} and \emph{Acipenser} and loci $\geq$ 50 bp, and we used this subset to estimate a species tree with CloudForest (\url{https://github.com/ngcrawford/CloudForest}), a parallel implementation of a workflow combining substitution model selection (identical to MrAIC 1.4.4 \cite{Nylander 2004}) and genetree estimation using PhyML \cite{Guindon2010}.  We estimated the species tree by summarizing gene trees using STAR \cite{Liu2009a, Liu2009b, Liu2010}.  To assess confidence in the resulting species tree, we used CloudForest to generate 1000, multi-locus, non-parametric bootstrap replicates by resampling nucleotides within loci as well as resampling loci within the data set \cite{Seo2008}, we summarized bootstrap replicates using STAR, and we reconciled bootstrap replicates with the species tree using RAxML.

\subsection*{Divergence Time Estimation}
We used a set of 14 calibration points (Appendix: Fossil calibrations) from previous timetree studies \cite{Hurley2007, Santini2009} to date several key splits on the tree ranging in age from the Givetian (392 Ma) to the early Rupelian (32 Ma).  We used the RAxML topology (Fig. \ref{fig:raxml-tree}) plus fossil-derived minimum and maximum age constraints to infer divergence times in BEAST v1.72 \cite{Drummond2007} assuming a GTR model of sequence evolution with gamma-distributed rate variation. Preliminary analyses with the full UCE  data (all 491 loci) showed poor mixing under a wide range of fossil constraint parameterizations, and all models incorporating the full UCE data set failed to reach convergence after 300 million generations. To estimate divergence times for these taxa from a large data set, we randomly sampled 50 UCE loci, with replacement, to create five different matrices for BEAST analyses. We analyzed the resulting matrices using independent runs of BEAST. MCMC runs of these data sets mixed much better and showed strong evidence of convergence, having ESS values $>$ 200 for nearly all parameters after 50 million generations. However, the parameters for the mean and variance of the lognormally distributed rates mixed much more slowly with autocorrelation times of $\geq 5*10^5$ generations and yielding ESS of $\approx$ 50 - 100 across the replicates. 

\section*{Results and Discussion}

\subsection*{Probe design, UCE enrichment, and sequencing}
We located 500 UCEs shared among all actinopterygian fishes, and we designed a set of 2,000 capture probes targeting each of these loci (4X coverage). Following enrichment and sequencing, we obtained an average of 2,819,047 reads per species, which we assembled into an average of 665 contigs having an average length of 457 bp (Table \ref{tab:read-stats}). After removing contigs that matched no UCEs and UCE loci that matched multiple contigs, we enriched an average of 332 (50\%) unique contigs matching UCE loci from each species. Average sequencing depth across UCE loci was 498X. We integrated extant genomic data from several fish species to this group of unique UCE contigs, and we constructed 491 alignments (average length: 305 bp, 95 CI: $\pm$16.0) comprised of 149,246 characters.  Each alignment contained an average of 21 target taxa (95 CI $\pm$ 0.4) after data trimming, and we used this incomplete data matrix for subsequent analyses with RAxML and MrBayes.  After removing loci having missing data for \emph{Polypterus} and \emph{Acipenser}, we input 136 alignments (41,731 characters; average length: 307 bp, 95 CI: $\pm$ 27.7) to CLOUDFOREST for model selection and subsequent species tree estimation using STAR.

\subsection*{A phylogenomic perspective on the basal radiation of ray-finned fishes}
\begin{figure}[ht!]
    \begin{center}
        \includegraphics[width=1.0\textwidth]{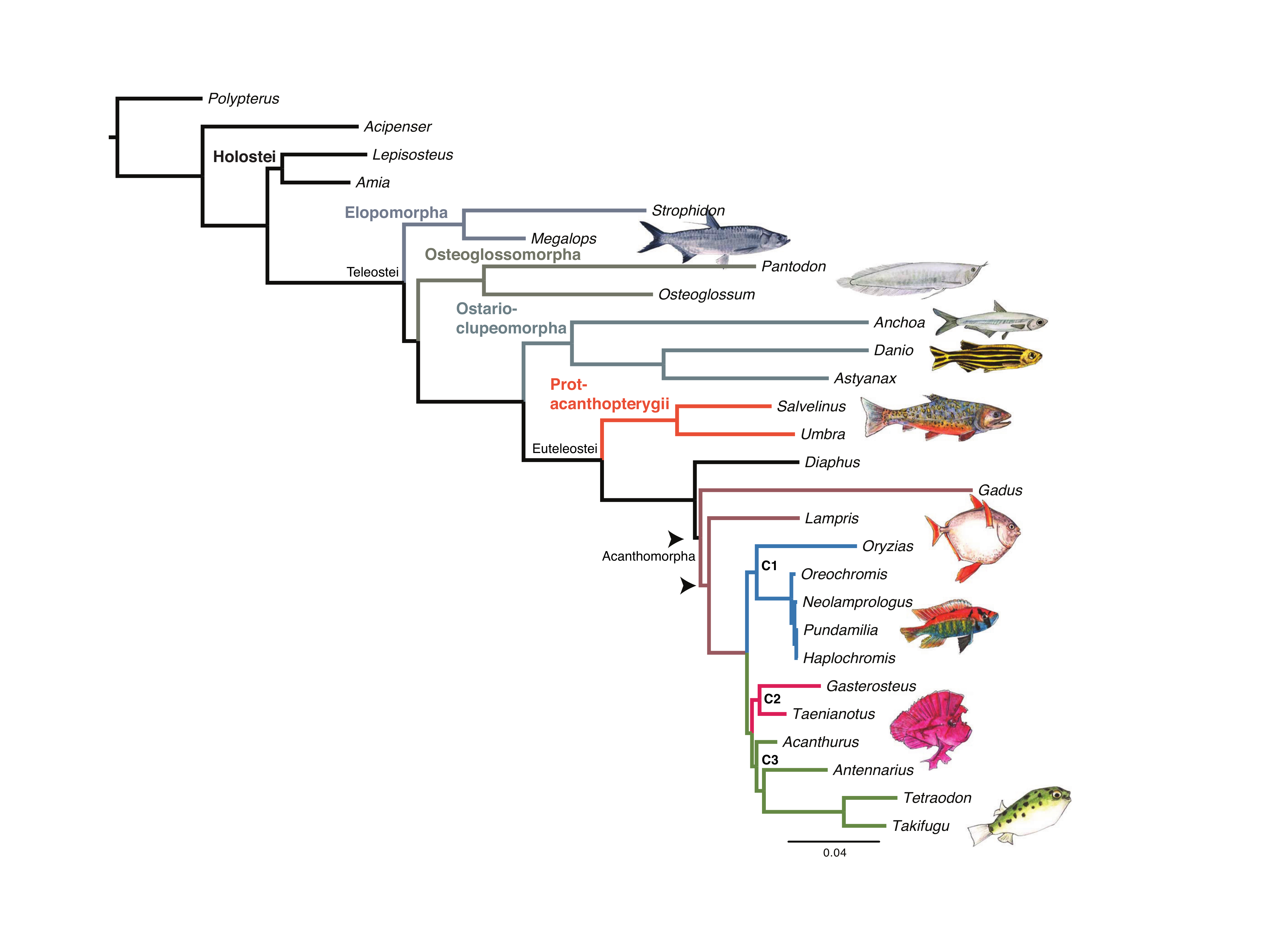}
    \caption{
        {\bf Maximum likelihood  phylogram of ray-finned fish relationships based upon UCE sequences.} All nodes except for two (indicated by arrows) supported by bootstrap proportions and Bayesian posterior probabilities $>$ 0.99. Our analysis supports a monophyletic Holostei and reveals the elopomorphs to be the earliest diverging lineage of teleosts. C1, C2, and C3 indicate clades within acanthomorphs consistent with other recent molecular studies (see Discussion). 
    }
    \label{fig:raxml-tree}
    \end{center}
\end{figure}

Maximum likelihood analysis produced a single, completely resolved topology wherein all but two nodes received high ($\ge$ 0.99) bootstrap proportions and Bayesian posterior probabilities (Fig. \ref{fig:raxml-tree}). This topology  provides new insight into several long-standing questions concerning the evolution of ray-finned fishes. Our analysis strongly supports the monophyly of the Holostei (\emph{Amia} + \emph{Lepisoteus}). This clade is historically controversial because morphological studies alternatively support \cite{Nelson1969, Jessen1972, Grande2010} and refute \cite{Olsen1984, Patterson1973} the monophyly of this group, while recent molecular studies generally recover the relationship \cite{Lietal2008, Inoueetal2003, Near2012}. Additionally, our analyses do not support  prior findings of an ``ancient fish clade'' including the Holostei + Aciperseriformes as the sister group to the teleosts  \cite{Inoueetal2003, Venkatesh2001}. Rather, our results strongly suggest a traditional relationship in which these lineages form successive sister groups to the teleosts. 

Our phylogenomic data provide strong evidence for the placement of elopomorphs as the sister group to all other teleosts and osteoglossomorphs and ostarioclupeomorphs as successive sister lineages to the euteleosts (Fig. \ref{fig:raxml-tree}). Our maximum likelihood topology is strongly incongruent with mitogenomic studies \cite{Inoue2003,  Inoue2005} but consistent with both a recent analysis of multiple nuclear genes \cite{Near2012} and some of the earliest morphological analyses of the group \cite{Arratia2001, Arratia2010, Arratia2004, Cloutier2004}. Within euteleosts, our results are congruent with recent molecular studies  \cite{Zou2012, Near2012, Li2008} in placing esociforms as the sister to salmoniforms rather than any neoteleost lineages. 

Within acanthomorphs, the largest clade of euteleosts, UCEs recover several intriguing clades that agree with results from recent molecular phylogenetic studies. These include the African cichlids + medaka (Clade C1, Fig. \ref{fig:raxml-tree}), corresponding to an expanded clade of atherinomorphs suggested by recent studies \cite{Mabuchi2007, Wainwright2012, Zou2012}; a clade of gasterosteiforms (stickleback) and scorpaeniforms (\emph{Taenionotus}) that is congruent with recent molecular and morphological studies \cite{Smith2006,  Smith2007, Zou2012}; and a clade including surgeonfish, frogfishes, and pufferfishes (acanthuroids, lophiiforms, and tetraodontiforms) corresponding to acanthomorph clade ``N'' of Dettai and Lecointre \cite{Dettai2005, Near2012}. Together with the estimated topology, our timetree suggests that UCEs provide sufficient phylogenetic signal to resolve divergences within fishes that are as young as 5 Ma and as old as 400 Ma.

The STAR topology was less resolved than topologies based upon analysis of the concatenated data set (Fig. \ref{fig:star-gt-st}) but recovered largely congruent relationships including a monophyletic Holostei as the sister to other actinopterygians, monophyly of elopomorphs, osteoglossomorphs, ostarioclupeomorphs, and euteleosts, and a successive sister group relationship between ostarioclupeomorphs, \emph{Salvelinus} + \emph{Umbra}, and all remaining euteleosts. The species tree switched the position of the Gadiformes, represented by cod (\emph{Gadus}) and Myctophiformes, represented by \emph{Diaphus}. This resolution is not congruent with results from Near \emph{et al.,} \cite{Near2012} but has been suggested before in other molecular studies \cite{Li2008, Meynard2012}.

\subsection*{A phylogenomic timescale for the radiation of ancient fish lineages}
A general expectation during Bayesian analysis is that runs will converge more quickly on the true posterior distribution with increasing data.  However, in our divergence time analysis, we were only able to obtain acceptable mixing in BEAST when analyzing subsamples of the data set; analyses of all 491 loci did not converge despite a diversity of partitioning and calibration strategies combined with long MCMC analyses. This was surprising given the relatively modest number of taxa included in this study and because we fully constrained the topology. Discordant fossil calibrations \cite{Near2005} could potentially underlie poor mixing behavior. However, alternative fossil calibration strategies that enforced subsets of the 14 calibrations revealed that even a single constraint on the age of the root node resulted in analyses with poor mixing behavior. We observed additional evidence that fossil conflict was not the cause of poor mixing because subsampled analyses including all 14 constraints converged and produced nearly identical credible intervals for all nodes (Fig. \ref{fig:div-estimates}). One possible barrier to good mixing during analyses of the full data set in BEAST may be that the proposal mechanisms and/or tuning parameters for branch lengths are not suited for alignments that contain both very quickly (UCE flanks) and very slowly (UCE core regions) evolving sites. With a large number of sites, even a relatively small change in branch length could result in a large difference in likelihood, resulting in a high rejection rate of branch length proposals. This explanation is consistent with the extremely low acceptance rates we observed in our analyses ( $<$ 5\% for  branch rate and node height proposals). The recovery of similar age estimates across UCE subsamples (Fig. \ref{fig:div-estimates}) suggests that the estimated time scale for ray-finned fish diversification adequately reflects the signal for divergences contained within the full UCE data set.
 
Our divergence time analyses suggest that the extant radiation of ray-finned fishes began $\approx$ 420 Ma near the end of the Silurian. Divergences amongst the actinopterians, neopterygians, and holosteans (Fig. \ref{fig:div-estimates}, nodes 1-4) are very ancient and span the late Devonian to the early Permian. Extant teleosts (Fig. \ref{fig:div-estimates}, node 5) can trace their origin back to $\approx$ 270 Ma during the late Permian with primary divergences amongst the elopomorphs, osteoglossomorphs, ostarioclupeomorphs, and euteleosts occurring before the end of the Triassic. The credible intervals surrounding the age of teleosts are wide, however, and teleosts may have originated as early as 309 Ma during the upper Carboniferous to as late as 226 Ma during the mid-Triassic. We recover a basal split between \emph{Salvelinus} $+$ \emph{Umbra}, representing two protancanthopterygian lineages, and the rest of the euteleosts, at 170 Ma during the Jurassic (Fig. \ref{fig:div-estimates}, node 9).  Extant acanthomorphs, which comprise the bulk of teleost diversity including over 16,000 species and 300 families, trace their origins to the Cretaceous, $\approx$ 124 Ma (Fig. \ref{fig:div-estimates}, node 14). We report ages for all splits in Table \ref{tab:divtimes}.

Our analysis represents the first timescale for ray-finned fishes derived from widespread sampling of the nuclear genome. Previous divergence time analyses based upon fossils, mitochondrial DNA, and nuclear DNA have produced conflicting timescales for ray-finned fish diversification. For example, the earliest known fossil teleosts are elopomorphs and ostariophysans, dating to the Late Jurassic \cite{Hurley2007, Santini2009}. In contrast mitogenomic data consistently suggest Paleozoic divergences, usually into the early Permian or late Carboniferous (e.g., \cite{Johnson2012}). Some divergence time studies based upon nuclear genes have posited Late Triassic to Early Jurassic divergences (173-214 Ma \cite{Santini2009, Hurley2007}) for the teleosts, while more recent studies that include additional nuclear loci and an improved set of fossil calibrations push the age of crown teleosts to 307 Ma at the end of the Carboniferous \cite{Near2012}. Our UCE-derived date for the divergence of teleosts ($\approx$ 270 Ma) is much older than an earlier estimate based upon \emph{RAG1} sequence \cite{Santini2009} and largely overlaps with the age estimate derived by Near et al \cite{Near2012}. Other splits within our UCE timescale generally correspond to those derived by Near et al \cite{Near2012} (Fig. \ref{fig:compare-dates}). One major exception to this observation is divergence of the euteleosts which UCE data estimate to be $\approx$ 171 Ma (95\% credible interval, 152-194 Ma) while traditional nuclear loci estimate an age of $\approx$ 229 Ma (95\% credible interval 220-259 Ma) \cite{Near2012}. One possible cause for this difference is age estimates is that we have not sampled some of the earliest diverging eutelost lineages in our study including \emph{Lepidogalaxias}, resulting in an underestimate of the true euteleost crown age. Our molecular timescale provides additional evidence for a relatively ancient origin of teleosts and further highlights the apparent gap in the fossil record between stem and crown members of this clade \cite{Hurley2007, Near2012}.  

\subsection*{Conclusions} Sequence capture of regions anchored by UCEs offers a powerful and efficient means of generating massive genomic data sets capable of resolving phylogenetic relationships at both deep and shallow scales in non-model organisms. Our UCE-based approach offers several advantages over previous studies that should contribute to the reliability of our topology.  These benefits include efficient sampling of sequence data across individual genomes and among divergent taxa, collection of data from an order of magnitude more loci than studies based upon traditionally-used genetic markers and almost twice as many loci as transcriptome-based genomic studies \cite{Zou2012}, validity of the UCE probe set across bony fishes spanning 350 Ma of evolutionary history, and utility of the UCE enrichment approach with tissues collected from museum specimens.  Additionally, these data illustrate that biologists can use UCE-based genetic markers to reconstruct the phylogeny of taxa other than amniotes, supporting the observation that UCE-based markers are a universal source of phylogenetically informative characters \cite{Faircloth2012, McCormack2012a}.

\section*{Acknowledgments}
National Science Foundation grants DEB-6861953 and DEB-6701648 (to MEA) and DEB-1242260 (to BCF) provided partial support for this work. Funds from an Amazon Web Services education grant (to BCF) supported computational portions of this work. We thank Bryan Carstens, Scott Herke, and the LSU Genomics Facility for help with Illumina sequencing.  We thank Travis Glenn for helpful discussion of several laboratory methods, and we thank John Huelsenbeck, Brian Moore, and John McCormack for helpful discussions relative to phylogenetic analyses. Marc Suchard provided helpful advice about BEAST analyses. Amisha Gadani provided illustrations. We obtained tissues used in this study on loan or as gifts from Peter Wainwright (UC Davis), Rita Metha (UC Santa Cruz), Mark Westneat (Field Museum), Eric Hilton and Patrick McGrath (Virginia Institute of Marine Sciences), David Jacobs and Ryan Ellingson (UCLA), H.J. Walker and Phil Hastings (Scripps Institute of Oceanography), Anindo Choudhury (St. Norbert College).

\section*{Author Contributions}
MEA, FS, LS, and BCF designed the study.  BCF identified conserved regions and designed capture probes. LS and BCF extracted DNA, prepared DNA libraries, and performed DNA sequence enrichments.  BCF processed sequence data, assembled sequence data, and created data sets for subsequent analysis.  BCF performed gene tree analyses.  BCF and MEA performed concatenated analyses.  MEA and FS performed dating analyses.  MEA, BCF, and FS wrote the manuscript.  All authors discussed results and commented on the manuscript.

\section*{Data Availability}
Contigs assembled from raw read data are available from Genbank (Accession \#s: XXX - YYY) [{\bf{We are in the process of submitting assembled genomic data to Genbank for archiving, which takes a bit of time given the total amount of data we are submitting}}].  Probe data, assembled contigs, alignments, and data sets we used for analysis are temporarily available from \url{http://fish-phylogenomics.s3-website-us-east-1.amazonaws.com/} (we do not log accesses of this website).  Upon acceptance of this manuscript, we will make all assembled contigs, capture probes, alignments, and datasets available from Dryad (\url{http://datadryad.org}).  Protocols for library preparation and UCE enrichment are available under Creative Commons license from \url{http://ultraconserved.org}.  Software used for the analysis of raw sequence data are available under BSD-style license from \url{https://github.com/faircloth-lab/splitaake}, \url{https://github.com/faircloth-lab/illumiprocessor}, and \url{https://github.com/faircloth-lab/phyluce}.

\newpage

\bibliography{alfaro-et-al-2012}

\newpage

\appendix

\section*{Description of timetree calibration points}

\begin{figure}[ht!]
    \begin{center}
        \includegraphics[width=1.0\textwidth]{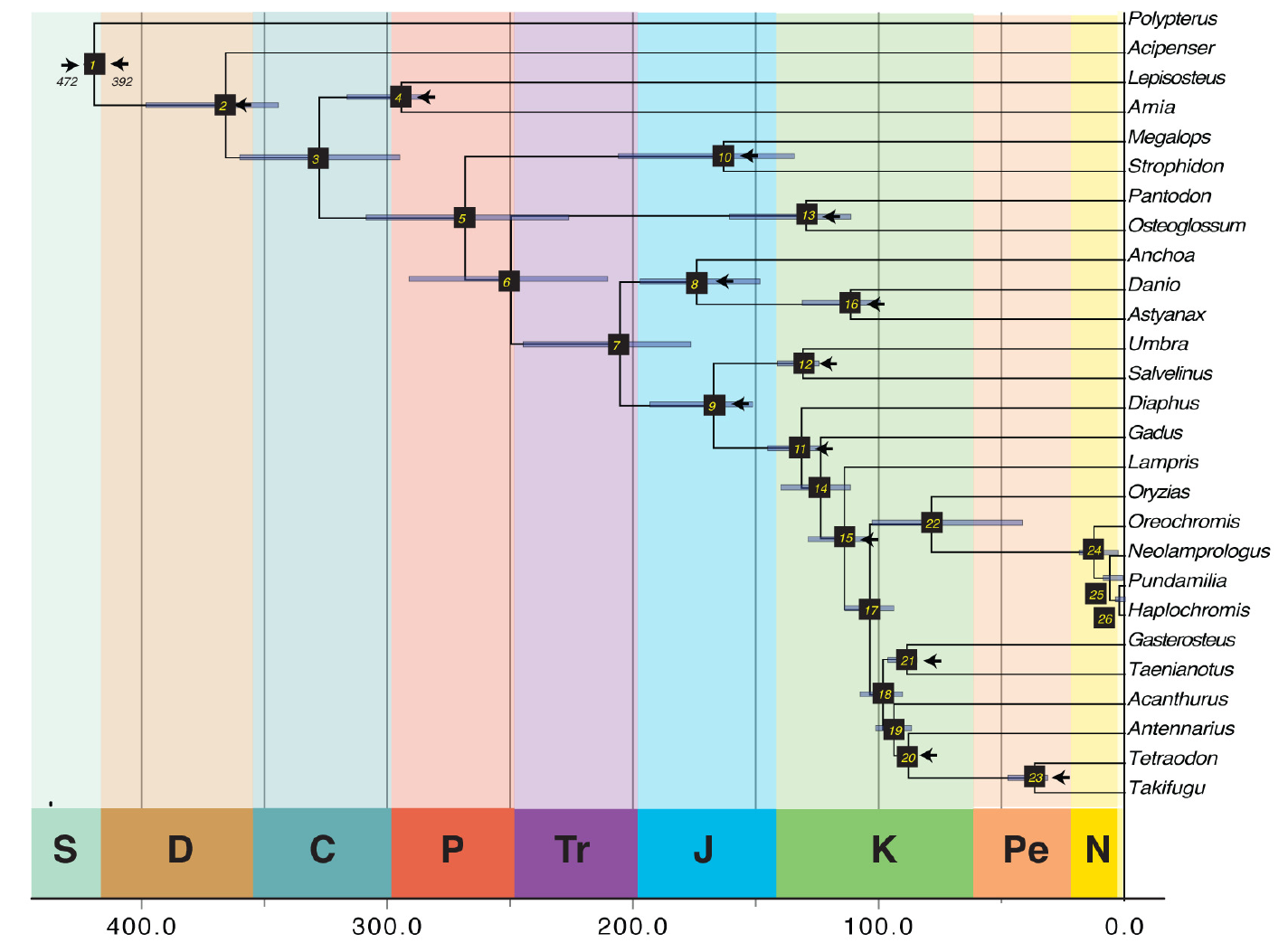}
    \end{center}
    \caption{
        {\bf UCE-based timescale of ray-finned fish diversification.} Arrows indicate fossil-calibrated nodes (described in Appendix 1). Numbers refer to splits in Fig. \ref{tab:divtimes}.
    }
    \label{fig:timescale}
\end{figure}

We assigned fossil calibrations to 14 nodes of the maximum likelihood phylogeny (Fig. \ref{fig:timescale}) to enable divergence time estimation using BEAST v1.72.  Below, we justify the bounds for each calibration point.

\begin{description}

\item[MRCA of Actinopterygii] \hfill \\
{\bf{Fig. \ref{fig:timescale}, node 1; uniform calibration = 392-472 Ma}}\\
We used the stegotrachelids fossils from the Givetian/Eifelian boundary ($\approx$ 392 Ma) used by \cite{Hurley2007} to date the minimum age of the root and assumed an upper bound of 472 Ma based upon the minimum age for the split between acanthodians and all other bony fishes \cite{Janvier 1996}.

\item[MRCA of Actinopteri  ] \hfill \\
{\bf{Fig. \ref{fig:timescale}, node 2, exponential calibration lower bound = 345 Ma, upper 95\% = 392 Ma}}\\
The oldest fossil belonging to this clade is \emph{Cosmoptycius} from the Tournasian (Carboniferous, 359-345 Ma)\cite{Hurley2007}. We used the stegotrachelids fossils from the Givetian/Eifelian boundary ($\approx$ 392 Ma) used by \cite{Hurley2007} to date the crown actinopterygians to set the upper bound.

\item[MRCA of Holostei] \hfill \\
{\bf{Fig. \ref{fig:timescale}, node 3; exponential calibration lower bound = 284 Ma, upper 95\% = 345 Ma}}\\
The oldest fossil assigned to this clade is the neopterygian \emph{Brachydegma caelatum}, from the Artkinsian (early Permian, 284 Ma) \cite{Hurley2007}. We used \emph{Cosmoptychius} from the Tournasian (Carboniferous, 359-345 Ma) to set the upper bound \cite{Hurley2007}.

\item[MRCA of Teleostei (Fig. \ref{fig:timescale}, node 5; unconstrained)] \hfill \\
The oldest crown teleost is \emph{Anaethalion}, from the late Kimmeridgian (Jurassic, 152 Ma) \cite{Santini2009}. However, due to the appearance of fossils from several major teleost groups (e.g., the ostarioclupeomorph \emph{Tischlingerichthys viohli}, the euteleost \emph{Leptolepides sprattiformis}) in deposits of the same age it seems likely that teleosts arose considerably earlier. We left this node unconstrained in our analysis to obtain an estimate based upon other actionpterygian calibrations and the UCE sequence data. 
 
\item[MRCA of Osteoglossomorpha] \hfill \\
{\bf{Fig. \ref{fig:timescale}, node 13; lower bound = 112 Ma, upper 95\% = 225 Ma}}\\
The oldest taxon assigned to our osteoglossomorph clade (which only includes a subset of the major osteoglossomorph lineages) is the arapamid \emph{Laeliichthys ancestralis} from the Aptian Areado Formation of Brazil (Cretaceous, 112 Ma) \cite{Patterson1993}. We used the Hiodontidae \emph{Yanbiania wangqingica} from the Barremian (Cretaceous, 130 Ma) to set the upper boundary \cite{Santini2009}.

\item[MRCA of Ostarioclupeomorpha] \hfill \\
{\bf{Fig. \ref{fig:timescale}, node 8; lower bound = 149 Ma, upper 95\% = 225 Ma}}\\
The oldest taxon assigned to this clade is \emph{Tischlingerichthys viohli}, from the upper Tithonian (Jurassic, 149 Ma) \cite{Santini2009}. We used the stem teleost \emph{Pholidophoretes salvus} (Pholidophoridae), from the early Carnian/Julian (Triassic, 228-225 Ma) to set the  upper bound \cite{Santini2009}. 

\item[MRCA of Euteleostei] \hfill \\
{\bf{Fig. \ref{fig:timescale}, node 9, lower bound  = 152 Ma, upper 95\% = 225 Ma}} \\
The oldest taxon assigned to this clade is \emph{Leptolepides sprattiformis} (Orthogonikleithridae) from the late Kimmeridgian (Jurassic, 152 Ma) \cite{Santini2009}. We used the stem teleost \emph{Pholidophoretes salvus} (Pholidophoridae), from the early Carnian/Julian (Triassic, 228-225 Ma) to set the upper bound \cite{Santini2009}. 

\item[MRCA of Elopomorpha] \hfill \\
{\bf{Fig. \ref{fig:timescale}, node 10, lower bound = 135 Ma, upper 95\% = 225 Ma}}\\
The oldest taxon assigned to this clade is the albulid \emph{Albuloideorum ventralis}, from the early Hauterivian (Jurassic/Cretaceous border, 135 Ma) \cite{Santini2009}. The oldest stem elopomorph is \emph{Anaethalion}, from the late Kimmeridgian (Jurassic, 152 Ma) \cite{Santini2009}. 

\item[Characiformes vs Cypriniformes] \hfill \\
{\bf{Fig. \ref{fig:timescale}, node 16, lower bound  = 100 Ma, upper 95\% = 149 Ma}}\\
The oldest taxon assigned to this clade is the characiform \emph{Santanichthys diasii} from the Albian (Cretaceous, 112-100 Ma) \cite{Santini2009}. We used \emph{Tischlingerichthys viohli}, from the upper Tithonian (Jurassic, 149 Ma) to set the upper bound \cite{Santini2009}. Our prior assumed 100 My as the minimum age, and 149 Ma for the upper bound.

\item[Salmoniformes vs Esociformes] \hfill \\
{\bf{Fig. \ref{fig:timescale}, node 12, lower bound  = 125 Ma, upper 95\% = 152 Ma}}\\
The oldest taxon assigned to this clade is \emph{Helgolandichthys schmidi}, from the early Aptian (early Cretaceous, 125 Ma) \cite{Santini2009}. We used \emph{Leptolepides sprattiformis} (Orthogonikleithridae) from the late Kimmeridgian (Jurassic, 152 Ma) to set the upper bound \cite{Santini2009}. 

\item[MRCA of Ctenosquamata] \hfill \\
{\bf{Fig. \ref{fig:timescale}, node 11, lower bound  = 122 Ma, upper 95\% = 152 Ma}}\\
The otoliths assigned to \emph{Acanthomorphorum forcallensis} from the Aptian (Early Cretaceous, 124-122 Ma) represents the oldest fossils assigned to this clade \cite{Santini2009}.We used \emph{Leptolepides sprattiformis} (Orthogonikleithridae) from the late Kimmeridgian (Jurassic, 152 Ma) to set the upper bound \cite{Santini2009}.

\item[MRCA of Acanthopterygii] \hfill \\
{\bf{Fig. \ref{fig:timescale}, node 15, lower bound  = 99 Ma, upper 95\% = 122 Ma}}\\
The oldest taxa assigned to this clade are various Beryciformes (e.g., \emph{Hoplopteryx}, \emph{Trachichthyoides}) from the Cenomanian (Late Cretaceous, 99 Ma) \cite{Santini2009}. We used the fossil otoliths assigned to \emph{"Acanthomorphorum" forcallensis} from the Aptian (Early Cretaceous, 124-122 Ma) to set the upper bound \cite{Santini2009}. 

\item[\emph{Gasterosteus} vs \emph{Taenianotus}] \hfill \\
{\bf{Fig. \ref{fig:timescale}, node 21, lower bound  = 85 Ma, upper 95\% = 122 Ma}}
We used the oldest gasterosteiform, \emph{Gasterorhamphus zuppichinii} from the Santonian (84-85 Ma) to date the minimum age of this split \cite{Patterson1993}. We used the fossil otoliths assigned to \emph{"Acanthomorphorum" forcallensis} from the Aptian (Early Cretaceous, 124-122 Ma) to set the upper bound \cite{Santini2009}. 
 
\item[Lophiiforms vs tetraodontiforms] \hfill \\
{\bf{Fig. \ref{fig:timescale}, node 20, lower bound  = 85 Ma, upper 95\% = 122 Ma}}\\
The oldest taxon assigned to this clade is the stem tetraodontiform \emph{Cretatriacanthus guidottii} from the Santonian of Nardo (Italy). We chose this taxon to date the minimum age rather than other, older, stem tetraodontiformes because preliminary re-examination of the relationships of extant and fossil tetraodontiforms (Santini, unpublished) casts doubt on their phylogenetic affinities. We used the fossil otoliths assigned to \emph{"Acanthomorphorum" forcallensis} from the Aptian (Early Cretaceous, 124-122 Ma) to set the upper bound \cite{Santini2009}. 

\item[MRCA of Tetraodontidae] \hfill \\
{\bf{Fig. \ref{fig:timescale}, node 23, lower bound  = 32 Ma, upper 95\% = 50 Ma}}\\
The oldest taxon assigned to this clade is \emph{Archaeotetraodon winterbottomi} from the Oligocene of Caucasus (32-35 Ma) \cite{Santini2009}. We used the stem tetraodontid \emph{Eotetraodon pygmaeus} from the Ypresian (middle Eocene, 50 Ma) to set the upper bound \cite{Santini2003}.
 
 \end{description}

\newpage
\nolinenumbers

\section*{Figure Legends}

\newpage

\setcounter{figure}{0}
\makeatletter 
\renewcommand{\thefigure}{S\@arabic\c@figure} 
\makeatother
%

\begin{figure}[h!]
	\begin{center}
    		\includegraphics[width=1.0\textwidth]{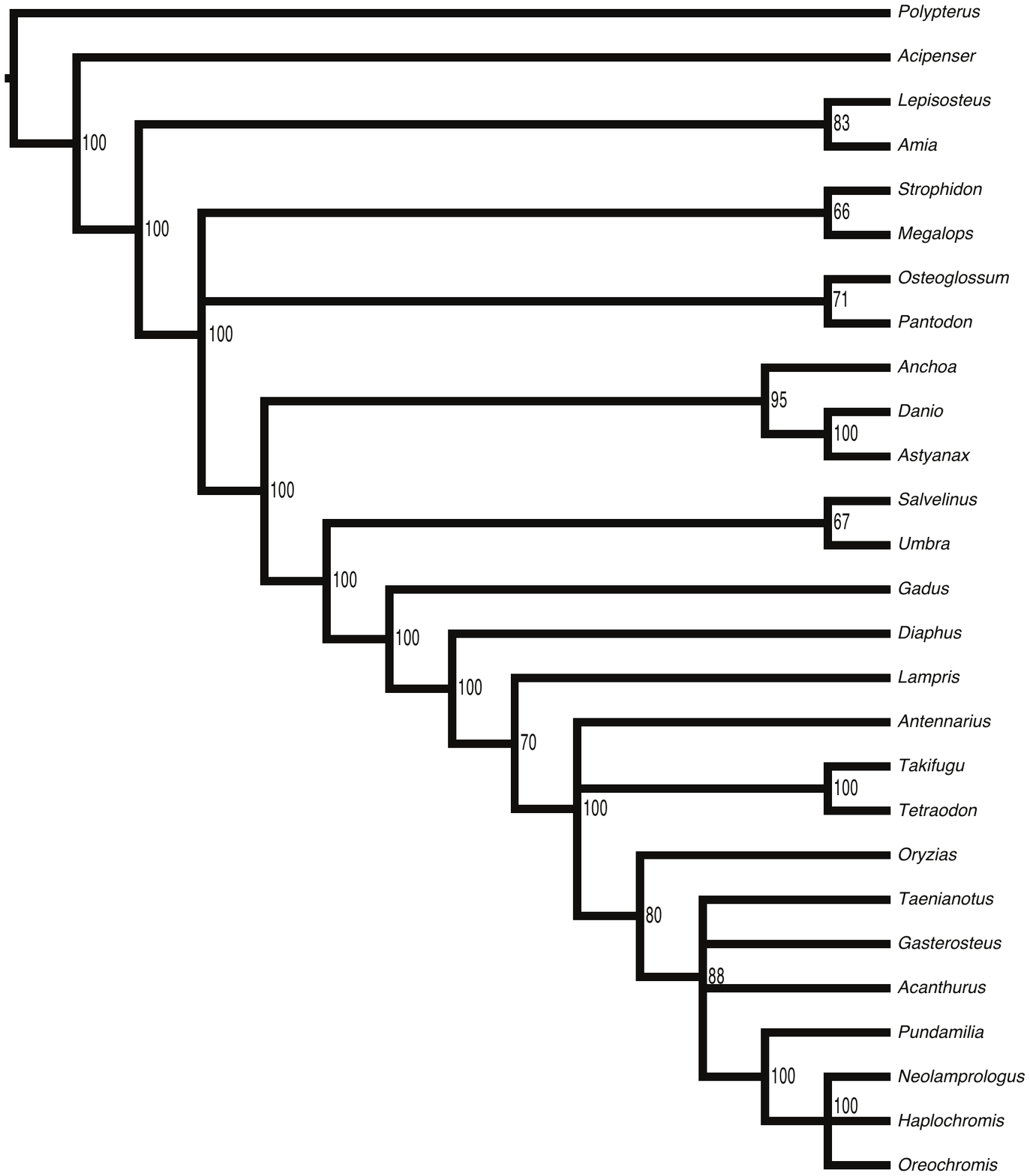}
	\caption{
	{\bf Species tree based upon STAR analysis.} Topology based upon analysis of all loci $\ge$ 50 base pairs that contained both \emph{Polypterus} and \emph{Acipenser} (N = 136). Node values indicate bootstrap proportion based upon 1000 replicates.  We collapsed nodes having $\leq$ 50\% bootstrap support.  
	}
	\label{fig:star-gt-st}
	\end{center}
\end{figure}

\newpage

\begin{figure}[h!]
	\begin{center}
    		\includegraphics[width=1.0\textwidth]{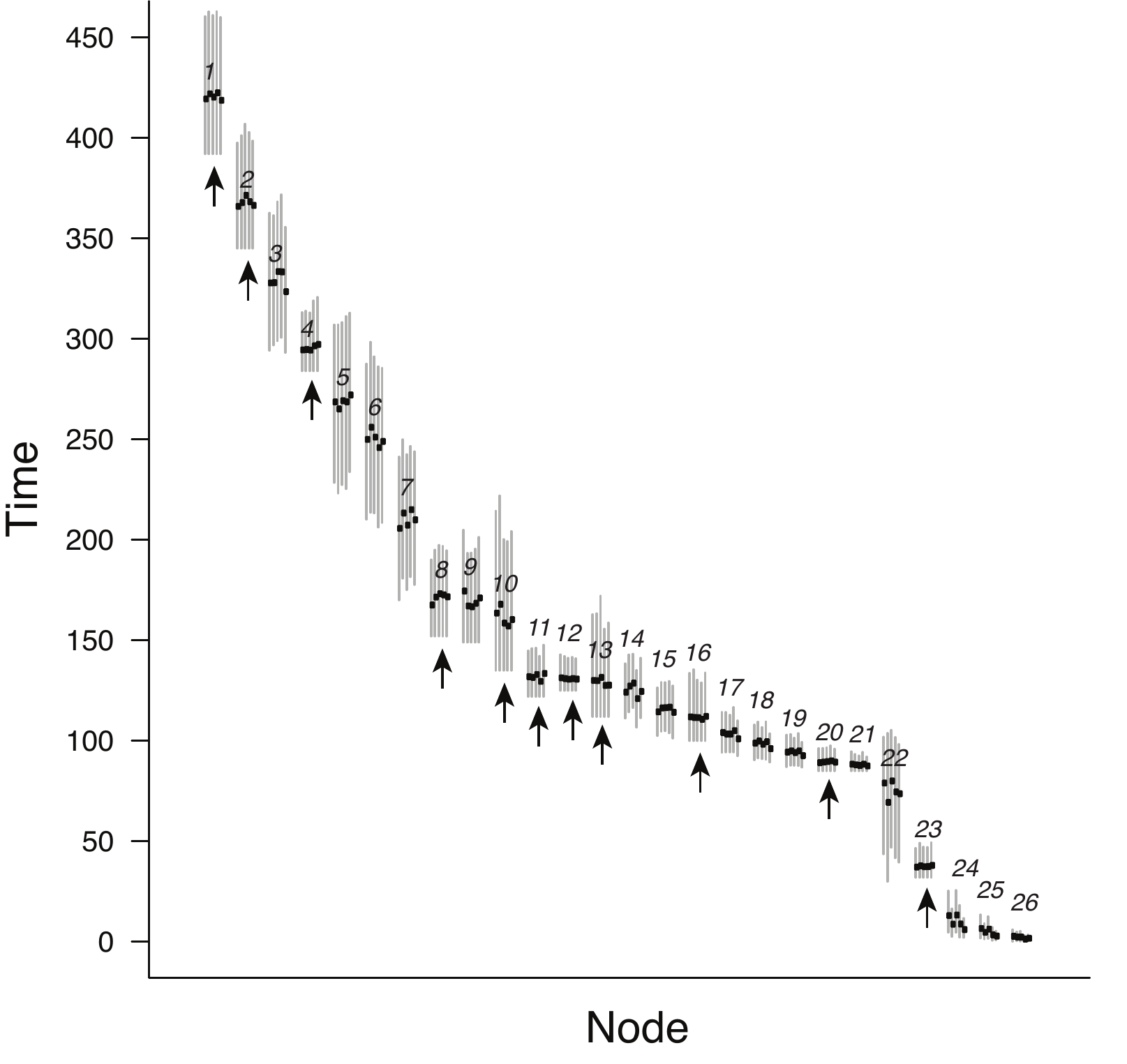}
	\caption{Mean and 95\% credible interval of divergence time estimates for subsamples of the UCE data set. Node numbers refer to labels in Fig. \ref{fig:timescale}. Arrows indicate fossil-constrained nodes. For each node, we show the mean and 95\% credible interval of five divergence time analyses.}\label{fig:div-estimates}
	\end{center}
\end{figure}

\clearpage

\begin{figure}[h!]
	\begin{center}
    		\includegraphics[width=1.0\textwidth]{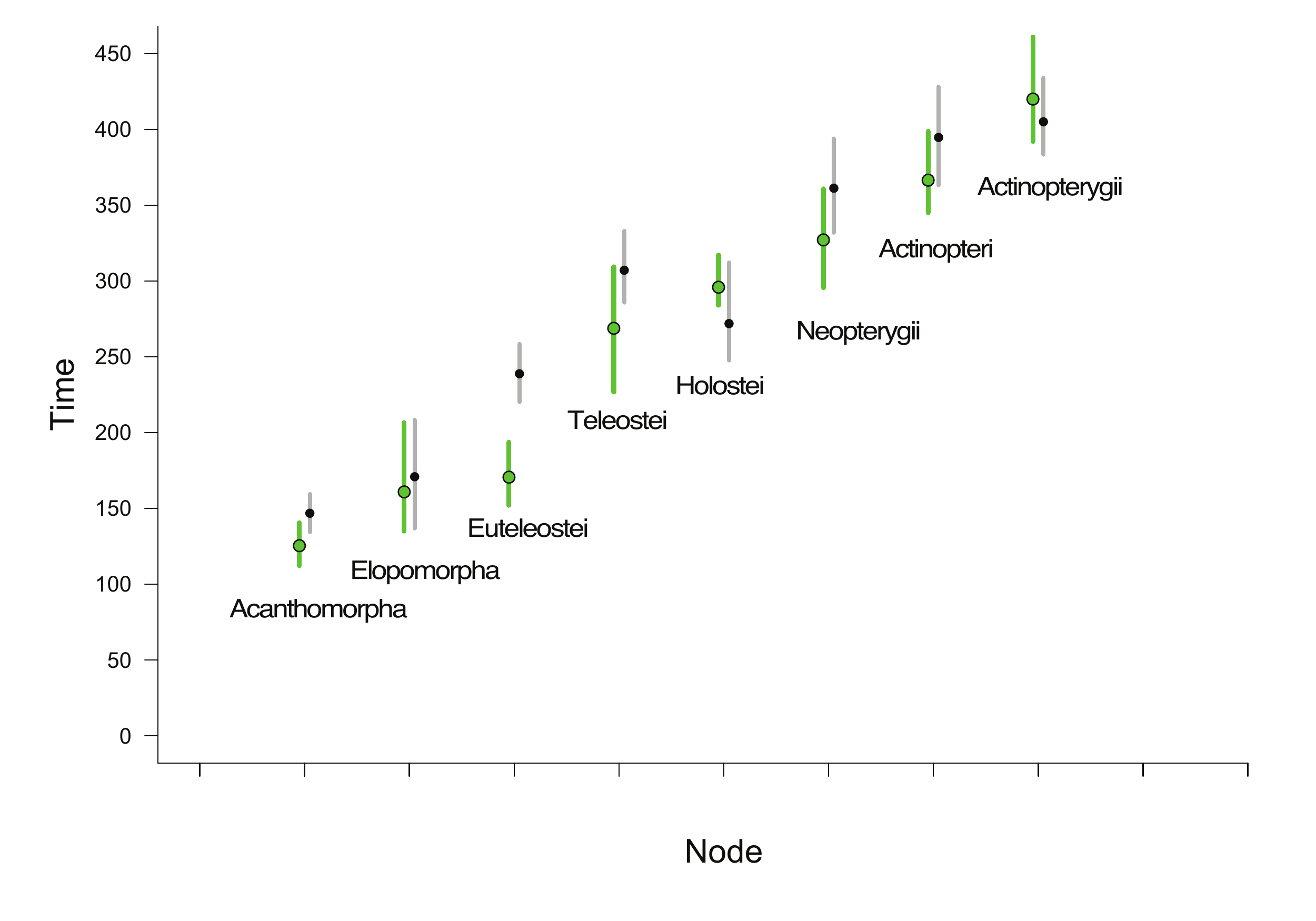}
	\caption{Comparison of ages derived from UCE data and nine nuclear genes \cite{Near2012}. UCE ages are in green. Circles indicate mean age and error bars indicate 95\% credible interval.} \label{fig:compare-dates}
	\end{center}
\end{figure}

\newpage

\section*{Tables}

\begin{table}[!ht]
\caption{Divergence times. Node numbers refer to Fig. \ref{fig:timescale}. We averaged mean and 95\% credible interval estimates (in Ma) across five subsampled divergence time analyses (see text).}
	\begin{center}
	\begin{tabular}{cccccc }
	\hline
Node	&	Clade	&	Split	&	Age 	&	Min	&	Max\\
1	&	Actinopterygii	&	\emph{Polypterus vs Takifugu}	&	420.0	&	392.0	&	461.0 \\
2	&	Actinopteri	&	\emph{Acipenser vs Takifugu}	&	366.5	&	345.0	&	398.8 \\
3	&	Neopterygii	&	\emph{Lepisosteus vs Takifugu}	&	327.1	&	295.5	&	360.8 \\
4	&	Holostei	&	\emph{Lepisosteus vs Amia}	&	295.9	&	284.0	&	317.1 \\
5	&	Teleostei	&	\emph{Megalops vs Takifugu}	&	268.8	&	226.8	&	309.4 \\
6	&	Osteoglossocephala	&	\emph{Pantodon vs Takifugu}	&	251.7	&	211.1	&	291.7 \\
7	&	Clupeocephala	&	\emph{Anchoa vs Takifugu}	&	210.9	&	177.1	&	245.3 \\
8	&	Ostarioclupeomorpha	&	\emph{Anchoa vs Astyanax}	&	169.5	&	149.0	&	197.8 \\
9	&	Euteleostei	&	\emph{Umbra vs Takifugu}	&	170.6	&	152.0	&	193.7 \\
10	&	Elopomorpha	&	\emph{Megalops vs Strophidon}	&	160.9	&	135.0	&	206.6 \\
11	&	Ctenosquamata	&	\emph{Diaphus vs Takifugu}	&	132.1	&	122.0	&	145.8 \\
12	&	protacanthopterygii	&	\emph{Umbra vs Salvelinus}	&	130.9	&	125.0	&	141.8 \\
13	&	Osteoglossiformes	&	\emph{Pantodon vs Osteoglossum}	&	128.9	&	112.0	&	161.5 \\
14	&	Acanthomorphs	&	\emph{Gadus vs Takifugu}	&	125.4	&	112.2	&	140.6 \\
15	&	--	&	\emph{Lampris vs Takifugu}	&	116.4	&	104.0	&	129.5 \\
16	&	Ostariophysans	&	\emph{Danio vs Astyanax}	&	111.4	&	100.0	&	131.9 \\
17	&	--	&	\emph{Oryzias vs Takifugu}	&	104.1	&	94.5	&	114.6 \\
18	&	--	&	\emph{Gasterosteus vs Takifugu}	&	99.1	&	91.0	&	108.2 \\
19	&	--	&	\emph{Acanthurus vs Takifugu}	&	94.1	&	87.4	&	102.0 \\
20	&	--	&	\emph{Gasterosteus vs Taenianotus}	&	89.7	&	85.0	&	97.0 \\
21	&	--	&	\emph{Antennarius vs Takifugu}	&	87.9	&	85.0	&	93.1 \\
22	&	--	&	\emph{Oryzias vs Haplochromis}	&	76.7	&	42.1	&	103.3 \\
23	&	Tetraodontidae	&	\emph{Tetraodon vs Takifugu}	&	37.5	&	32.0	&	48.0 \\
24	&	Cichlidae	&	\emph{Oreochromis vs Haplochromis}	&	9.9	&	3.2	&	19.2 \\
25	&	--	&	\emph{Neolamprologus vs Haplochromis}	&	4.7	&	1.3	&	9.4 \\
26	&	--	&	\emph{Pundamilia vs Haplochromis}	&	2.0	&	0.3	&	4.4 \\
	\hline
  	\end{tabular}
  	\label{tab:divtimes}
	\end{center}
\end{table}

\newpage

\begin{landscape}
\begin{table}[!ht]\footnotesize
\caption{Sequence read and assembly statistics for fish species used in this study.}
	\begin{centering}
	\begin{tabular}{| l | l | c | c | c | c | c | c | c | c | c | c | c |}
	\hline
	Scientific						&Common					& Number of		& Contigs			&	 Reads	 &UCE			& Reads in	&	Avg. 		&	Avg. 		&	Contigs 	&Reads \\
	name 						&name 						& trimmed 		& assembled		&	 in 		 &contigs			& UCE		&	size		&	coverage	&	on target	&on  \\
								&							& reads			&				&	 contigs	&				& contigs		&			&			&			& target \\
	\hline
	\emph{Umbra limi}				&	central mudminnow			&	2,727,071	&	1109				&	740,079	&		409		&	564,715	&	508.8	&	267.4	&	0.37	&	0.21	\\
	\emph{Diaphus theta}			&	California headlightfish		&	2,626,413	&	584				&	688,635	&		401		&	604,295	&	502.4	&	299.1	&	0.69	&	0.23	\\
	\emph{Antennarius striatus}		&	striated frogfish				&	3,724,320	&	474				&	2,462,193	&		418		&	2,310,186	&	649.7	&	850.2	&	0.88	&	0.62	\\
	\emph{Megalops} sp.			&	tarpon					&	2,771,805	&	786				&	650,577	&		247		&	231,314	&	485.4	&	191.5	&	0.31	&	0.08	\\
	\emph{Astyanax fasciatus}		&	banded astyanax			&	2,731,668	&	543				&	1,444,767	&		355		&	1,211,903	&	526.2	&	657.2	&	0.65	&	0.44	\\
	\emph{Acanthurus japonicus}		&	Japan surgeonfish			&	2,017,174	&	613				&	1,242,932	&		454		&	1,125,871	&	600.8	&	405.9	&	0.74	&	0.56	\\
	\emph{Amia calva}				&	bowfin					&	2,619,643	&	562				&	1,608,614	&		366		&	1,368,091	&	578.9	&	646		&	0.65	&	0.52	\\
	\emph{Lampris guttatus}			&	opah						&	2,472,439	&	486				&	1,350,852	&		418		&	1,237,650	&	568.7	&	520.2	&	0.86	&	0.50	\\
	\emph{Acipenser fulvescens}		&	lake sturgeon				&	3,083,152	&	577				&	1,129,829	&		167		&	467,414	&	426.9	&	665.4	&	0.29	&	0.15	\\
	\emph{Anchoa compressa}		&	deep body anchovy			&	2,617,717	&	533				&	783,323	&		287		&	625,862	&	448.6	&	479.2	&	0.54	&	0.24	\\
	\emph{Danio rerio}				&	zebrafish					&	2,777,132	&	518				&	1,367,065	&		382		&	1,166,020	&	463.4	&	657.1	&	0.74	&	0.42	\\
	\emph{Polypterus senegalus}		&	gray bichir				&	3,206,418	&	576				&	873,104	&		294		&	726,100	&	557.6	&	440		&	0.51	&	0.23	\\
	\emph{Pantodon buchholzi}		&	freshwater butterflyfish		&	3,329,691	&	466				&	2,058,929	&		272		&	1,399,286	&	550.4	&	930.5	&	0.58	&	0.42	\\
	\emph{Strophidon sathete}		&	slender giant moray			&	3,159,269	&	1007				&	448,390	&		277		&	246,758	&	510.6	&	172.9	&	0.28	&	0.08	\\
	\emph{Osteoglossum bicirrhosum}	&	silver arawana				&	2,735,138	&	643				&	1,565,346	&		276		&	813,175	&	467		&	623.9	&	0.43	&	0.30	\\
	\emph{Salvelinus fontinalis}		&	brook trout				&	2,466,696	&	1118				&	688,684	&		166		&	161,214	&	408.7	&	234.8	&	0.15	&	0.07	\\
	\emph{Taenianotus triacanthus}	&	leaf scorpionfish			&	3,245,453	&	712				&	1,423,244	&		447		&	1,252,564	&	652.5	&	431.4	&	0.63	&	0.39	\\
	\hline
  	\end{tabular}
	\end{centering}
  	\label{tab:read-stats}
\end{table}
\end{landscape}

\end{document}